\pdfoutput=1

\documentclass{webofc}
\usepackage[varg]{txfonts}

 \renewcommand{\baselinestretch}{0.978}

\usepackage{paracol} 

\usepackage{silence}
\usepackage{comment}

\usepackage{listings}
\usepackage[colorlinks=true, citecolor=blue, linkcolor=blue, urlcolor=blue, pdfborder={0 0 0}, breaklinks=true]{hyperref} 
\renewcommand{\url}[1]{\href{#1}{#1}}
\renewcommand{\doi}[1]{\url{https://doi.org/#1}}
\newcommand{\arxiv}[1]{\href{https://arxiv.org/abs/#1}{arxiv:#1}}

\usepackage{datetime2}

\usepackage{tikz}
\usepackage[compat=1.1.0]{tikz-feynman}
\tikzfeynmanset{warn luatex=false} 

\usepackage{multirow}

\usepackage[switch]{lineno}

\newcommand{\linenoDOC}{}
\newcommand{\linenoFIG}{}

\newcommand{\paperversion}{\hfill Version 2.0 (13 July 2021)}

\begin{document}
\linenoDOC

\vspace*{-2cm} 

\newcommand{\mgamc}{MG5aMC} 

\title{Design and engineering 
of a simplified workflow
execution for 
the \mgamc\ event 
generator on GPUs and vector CPUs}

\author{
\firstname{Andrea} \lastname{Valassi}
\inst{1}\fnsep
\thanks
{\email{andrea.valassi@cern.ch}
\hspace*{4.05cm} {\em\paperversion}
}
\and
\firstname{Stefan} \lastname{Roiser}
\inst{1}\fnsep
\and
\firstname{Olivier} \lastname{Mattelaer}
\inst{2}
\and
\firstname{Stephan} \lastname{Hageboeck}
\inst{1}
}
\institute{
CERN, IT-SC group,
Geneva, Switzerland
\and
Universit\'e Catholique de Louvain, Belgium
}

\abstract{
Physics event generators
are essential components
of the data analysis software chain
of high energy physics experiments,
and important consumers
of their CPU resources.
Improving the software performance
of these packages on modern
hardware architectures,
such as those deployed at HPC centers,
is essential in view of the upcoming 
HL-LHC physics programme.
In this paper,
we describe an ongoing activity
to reengineer
the Madgraph5\_aMC@NLO
physics event generator,
primarily to port it and allow 
its efficient execution on~GPUs,
but also 
to modernize it and
optimize its performance
on vector CPUs.
We describe the motivation,
engineering process
and software architecture design
of our developments,
as well as 
the current challenges and 
future directions
for this project.
This~paper 
is based on 
our submission to vCHEP2021 in March~2021,
complemented with a few 
preliminary results
that we presented 
during the conference.
Further details 
and updated results
will be given 
in later publications.
}

\maketitle

\newcommand{\Nvidia}{Nvidia}

\newcommand{\xvec}{\mathbf{x}}
\newcommand{\avec}{\mathbf{a}}
\newcommand{\fxvec}{f(\xvec)}
\newcommand{\xveci}{{\xvec_i}}
\newcommand{\aveci}{{\avec_i}}
\newcommand{\wi}{{w_i}}

\newcommand{\eplus}{$e^+$}
\newcommand{\eemumu}{\mbox{$e^+\!e^-\!\!\rightarrow\!\mu^+\!\mu^-$}}
\newcommand{\ggtt}{\mbox{$gg\!\rightarrow\! t\bar{t}$}}
\newcommand{\ggttg}{\mbox{$gg\!\rightarrow\! t\bar{t}g$}}
\newcommand{\ggttgg}{\mbox{$gg\!\rightarrow\! t\bar{t}gg$}}

\vspace*{-2mm}
\vspace*{-3mm}

\section{Introduction}
\label{sec_intro}
MadGraph5\_aMC@NLO~\cite{bib:mg5amc}
(in the following, \mgamc)
is a physics event generator software 
used in the data processing workflows
of High Energy Physics (HEP) experiments,
such as ATLAS and CMS
at CERN's Large Hadron Collider (LHC).
While the majority of CPU 
resources in the distributed 
computing environments
of the LHC experiments is spent
on detector simulation 
and event reconstruction workflows,
physics event generators 
are also large consumers of CPU time,
accounting for an estimated 
12\% of overall CPU budgets for ATLAS
and 5\% for CMS~\cite{bib:csbs}.
Their computational cost is 
also 
predicted to increase 
during the High-Luminosity phase of LHC (HL-LHC),
as more accurate theoretical predictions
are required for physics analyses.
Reengineering 
event generators
to improve their software performance
in view of the HL-LHC,
especially on modern hardware architectures
such as CPUs with many cores and wide vector registers,
is therefore essential~\cite{bib:csbs,bib:lhcctalk}.

With wider availability of graphics processing
units (GPUs) for computing,
the question also arises about
whether event generators like \mgamc\ may 
efficiently exploit 
these new architectures.
In recent years,
many high performance computing centers (HPCs) 
have deployed heterogeneous systems,
where the majority of the computing power
is provided by GPUs. 
These resources are currently
under-utilized by HEP experiments,
also because 
the port to GPUs 
of other HEP offline Grid workflows,
such as detector simulation,
has not yet been achieved.
Porting event generators to GPUs
would therefore be 
especially important.

The current production version of the \mgamc\ software
has been developed for CPUs only.
Previous work for porting 
some of its components to CUDA~\cite{bib:cuda} 
on \Nvidia\ GPUs
was done around 10 years ago~\cite{bib:heget1,bib:heget2,bib:heget3,bib:kan1,bib:kan2},
but unfortunately 
never reached production quality.
This effort has been restarted in early 2020 by
the authors of this paper
(including one of main authors 
of the \mgamc\ software), 
eventually 
as part of 
the activities of 
a larger team.
The project~\cite{bib:mg5-github-home} is a
collaboration between theorists,
experimentalists and software engineers,
which largely came about through
the activities of
the HSF Physics Event Generator WG~\cite{bib:hsfwg}.
This paper describes 
our software development process 
and engineering work 
for the port of \mgamc\ to GPUs,
and more specifically
on \Nvidia/CUDA,
and for efficiently exploiting vector CPUs 
by an improved C++ implementation.
Our colleagues in the larger
development team are also working
on a port to GPU architectures
from other hardware vendors,
notably through the use 
of abstraction layers,
but this work will eventually be described elsewhere.
While our development 
focuses on \mgamc,
we believe that many 
of the ideas and methods 
we use may also be useful 
for the software reengineering of
other event generators.

This paper is based on our 
March 2021 submission
to the vCHEP2021~\cite{bib:vchep} conference,
complemented with some 
preliminary results
that we presented~\cite{bib:vchep-talk}
there.
Its structure is the following.
We provide a brief description 
of the \mgamc\ code
in Sec.~\ref{sec_mgamc}.
An overview of~the 
engineering process is 
given in Sec.~\ref{sec_engproc}.
We describe 
the high-level 
software architecture design, 
some 
implementation details
and 
work in progress
in Sec.~\ref{sec_design}.
Our
preliminary results and 
future plans
are summarized 
in Sec.~\ref{sec_results}.
We give our interim conclusions
in~Sec.~\ref{sec_conclude}.
Further details 
and updated results
for this ongoing activity
will be presented 
in later 
publications.

\newcommand{\figfeyn}{
\begin{figure}[tb]
\begin{center}
\begin{tikzpicture}
\definecolor{forestgreenweb}{rgb}{0.13, 0.55, 0.13}
\newcommand{\fcol}{black} 
\newcommand{\fcolt}{\color{forestgreenweb}} 
\hspace*{-32mm}
\begin{feynman}[large]
\vertex[dot, blue] (eeg) {}; 
\vertex[\fcol, above left=of eeg] (em){\(e^{-}\)};
\vertex[\fcol, below left=of eeg] (ep){\(e^{+}\)};
\vertex[dot, red, right=of eeg, xshift=5mm] (mmg) {};
\vertex[\fcol, above right=of mmg] (mp){\(\mu^{+}\)};
\vertex[\fcol, below right=of mmg] (mm){\(\mu^{-}\)};
\diagram* {
(em) -- [\fcol, fermion, 
edge label = {\fcolt 1. \texttt{IXXXXX}},
edge label' = \raisebox{-6mm}{\color{black}\hspace*{-10mm}\Large(a)}] 
(eeg) -- [\fcol, fermion, 
edge label = {\fcolt 1. \texttt{OXXXXX}}] (ep),
(eeg) -- [photon, blue, edge label=\(\gamma\), 
edge label' = \raisebox{-3mm}
{\hspace*{-6mm}2. \texttt{FFV1P0\_3}}] (mmg),
(mp) -- [\fcol, fermion, 
edge label' = {\fcolt 1. \texttt{IXXXXX}},
edge label = \raisebox{-6mm}
{\color{red}\hspace*{-5mm}3. \texttt{FFV1\_0}}] 
(mmg) -- [\fcol, fermion, 
edge label' = {\fcolt 1. \texttt{OXXXXX}}] (mm)};
\end{feynman}
\hspace*{64mm}
\begin{feynman}[large]
\vertex[dot, blue] (eez) {}; 
\vertex[\fcol, above left=of eez] (em){\(e^{-}\)};
\vertex[\fcol, below left=of eez] (ep){\(e^{+}\)};
\vertex[dot, red, right=of eez, xshift=5mm] (mmz) {};
\vertex[\fcol, above right=of mmz] (mp){\(\mu^{+}\)};
\vertex[\fcol, below right=of mmz] (mm){\(\mu^{-}\)};
\diagram* {
(em) -- [fermion, \fcol, 
edge label = {\fcolt 1. \texttt{IXXXXX}},
edge label' = \raisebox{-6mm}{\color{black}\hspace*{-10mm}\Large(b)}] 
(eez) -- [fermion, \fcol, 
edge label = {\fcolt 1. \texttt{OXXXXX}}] (ep),
(eez) -- [boson, blue, edge label=\(Z\), 
edge label' = \raisebox{-3mm}
{\hspace*{-6mm}2. \texttt{FFV2\_4\_3}}] (mmz),
(mp) -- [fermion, \fcol, 
edge label' = {\fcolt 1. \texttt{IXXXXX}},
edge label = \raisebox{-6mm}
{\color{red}\hspace*{-5mm}3. \texttt{FFV2\_4\_0}}] 
(mmz) -- [fermion, \fcol, 
edge label' = {\fcolt 1. \texttt{OXXXXX}}] (mm)};
\end{feynman}
\end{tikzpicture}
\vspace*{-5mm}
\end{center}
\linenoFIG
\caption{
The two Feynman diagrams 
contributing to \eemumu at tree level:
(a) photon exchange 
and (b) Z exchange.
The routines called by \mgamc\ to 
compute the ME for each event
are also indicated.
First, the wavefunctions 
of the four external fermions
are computed
using \texttt{IXXXXX} and \texttt{OXXXXX},
given their momenta and helicities.
This is only done once,
as the result is the same for both diagrams.
Next, the wavefunctions of
the $\gamma$ and $Z$ propagators
in the two diagrams are computed 
from their coupling 
to the initial state electrons,
using \texttt{FFV1P0\_3} and \texttt{FFV2\_4\_3}.
Finally, the transition amplitudes
are computed
from the boson couplings 
to the final state muons
in the two diagrams,
using \texttt{FFV1\_0} and \texttt{FFV2\_4\_0}.
The two amplitudes are 
added to each other,
their sum is squared,
and the result is
averaged over all valid helicity combinations
to obtain the matrix element.
Each of the four fermions 
has two possible helicity states,
therefore there are in principle
sixteen helicity combinations;
however, as the $\gamma$ and $Z$
are vector bosons with spin 1,
only four of those 
are allowed.
Drawn with TikZ-Feynman~\cite{bib:tikzfey}.
}
\label{fig:fig1}
\vspace*{-5mm}
\end{figure}
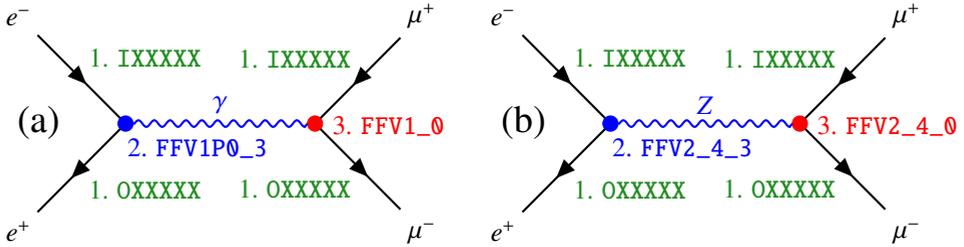
} 

\section{Brief description of the \mgamc\ software}
\label{sec_mgamc}

MadGraph5\_aMC@NLO (\mgamc)~\cite{bib:mg5amc}
is a computer program 
which aims at providing 
in a unified framework
all the elements necessary 
for the phenomenological study
of a wide spectrum of HEP collision processes,
both within the Standard Model (SM) and beyond (BSM).
In particular, \mgamc\ allows
the computation of cross sections
and generation of hard events
at tree level (LO) and next-to-leading order (NLO),
their matching~\cite{bib:mcnlo} 
to parton shower (PS) simulations,
and the merging of matched samples 
that differ by the number 
of jets (i.e.~gluons 
and light quarks)
in the final state.
Only a subset of 
these many features
are relevant to 
the work presented in this paper,
which is presently 
focusing on simple 
LO processes.

The general techniques 
used in \mgamc\ to integrate a LO 
partonic cross section, 
and to obtain a set of unweighted events from it, 
have been inherited from MadGraph5~\cite{bib:madgraph5}.
This strategy has then undergone
successive optimisations,
including some very recent ones~\cite{bib:kiran}.
The core of the process
is the calculation of 
tree-level invariant 
amplitudes and ``matrix elements'' (ME)
from the set of all relevant Feynman diagrams.
Since this depends 
on the collision process of interest,
\mgamc\ is constructed as
a Python meta-code, 
which is capable of automatically generating
the ME code
specific to the desired process.
The Lagrangian of the physics model 
can be expressed 
in a Python based format,
the Universal FeynRules Output (UFO)~\cite{bib:ufo}, 
thanks to various codes 
such as FeynRules~\cite{bib:feynrules},  
LanHep~\cite{bib:lanhep} 
and Sarah~\cite{bib:sarah}. 
Taking a given process and UFO as inputs, 
MG5aMC 
determines the set of relevant
Feynman diagrams 
and generates the code to evaluate 
the associated transition amplitude 
for an arbitrary phase space point. 
Thanks to the flexibility of Python, 
the code can be generated in 
various programming languages: 
Fortran is the default, but 
C++ and Python are supported as well.
The phase space points
where the ME calculation 
functions are evaluated
are determined by a Monte Carlo (MC) 
integrator, using a sampling algorithm
based on random number generation.
The integrator is the component that
drives the whole process of event generation,
cross section integration and event unweighting:
in the production version of \mgamc,
this is the Fortran package 
MadEvent, 
which uses an optimized 
phase space sampling algorithm
based on single diagram 
enhancement~\cite{bib:madevent,bib:kiran}.

\figfeyn

The calculation of matrix elements
from Feynman diagrams 
in \mgamc\ is based on 
helicity amplitudes~\cite{bib:kleiss,bib:hagzep1,bib:hagzep2,bib:aloha}
and QCD (Quantum Chromo Dynamics)
color-flow decomposition~\cite{bib:colorflow}.
With respect to 
trace techniques based on
completeness relations~\cite{bib:halzen}, 
the main advantage of 
this method
is that 
interference terms are 
handled by a simple sum over amplitudes
and, therefore, 
the computational complexity 
grows linearly 
with the number of diagrams instead of quadratically. 
In practice,
amplitudes 
(which are complex numbers)
for all diagrams
are computed numerically
for a given combination of helicities 
of the external~(initial and final state) particles,
and are then 
organized into 
subsets, each
corresponding to a given QCD
color flow (i.e.~a given
combination of colors) 
of the external particles. 
The amplitudes 
associated with different diagrams
for a 
color flow are summed
to give the so called dual amplitude
for that color flow.
The squared amplitude 
for a given helicity combination
is then computed as a quadratic form
on the vector of 
the dual amplitudes 
for different color flows,
using a pre-computed color matrix,
which is the same for all events.
Finally, 
the average of these squared amplitudes 
over all helicity combinations yields
the overall matrix element
for the 
collision event. 
In this last step, only 
a specific subset of 
helicity combinations
have a non-zero contribution to the ME,
as most combinations
are forbidden by the conservation of angular momentum;
to speed up the computation
for large samples of events,
\mgamc\ pre-determines 
which helicity combinations are allowed
by calculating their contributions 
on a small set of events. 
In the example of the \eemumu\ collision
shown in Fig.~\ref{fig:fig1},
sixteen different helicity combinations exist, as 
there are four external fermions
with two possible spin polarizations each,
but only four of those combinations 
are allowed by conservation laws.

The transition amplitude for a given process 
is computed in \mgamc\ using 
the helicity amplitude routines 
provided by the sub-program 
ALOHA~\cite{bib:aloha}.
Those routines are almost identical 
to the ones of the HELAS 
library~\cite{bib:helas1,bib:helas2}, 
which was used in MadGraph4.
While HELAS
functions are all hardcoded,
ALOHA generates 
most of the required~functions dynamically.
The main advantage of this approach,
adopted in the MadGraph5~\cite{bib:madgraph5} release,
is that it is not limited to 
predetermined physics models,
and can be easily extended.
This is also essential in the specific
work presented in this paper,
as it makes it easier to 
port and optimize the software
for new GPU and CPU architectures
directly in the \mgamc\ code base,
without relying on changes to external libraries.
A distinctive feature
of helicity amplitude methods
is that each particle in a Feynman diagram
is associated to a ``wavefunction''
that contains its spinorial representation:
this is a vector of complex numbers
whose dimensions depend 
on the properties of each particle
(essentially, on its spin).
For instance,
as in HELAS~\cite{bib:helas1}, 
the wavefunction for a fermion in ALOHA
is a vector of six complex numbers:
two for the chirality-left spinor,
two for the chirality-right spinor,
and two for its 4-momentum.
In the helicity wavefunctions formalism, 
the  computation of the ME code 
is performed thanks to three categories 
of routines used in a sequential way,
as shown in Fig.~\ref{fig:fig1}
(and later in Fig.~\ref{fig-mecompl}).
First, 
a wavefunction
is computed for 
each initial or final state 
particle in the Feynman diagram,
given its 4-momentum and helicity
using an ``external particle'' routine
(e.g.~\texttt{IXXXXX} for the ingoing 
and \texttt{OXXXXX} for the outgoing 
electrons and muons in Fig.~\ref{fig:fig1}).
Second, 
for each vertex in the Feynman diagram
which has one leg 
with no associated wavefunction,
this is computed 
using an ``internal particle'' 
routine
(e.g.~\texttt{FFV1P0\_3} for the 
off-shell propagator 
at the electron-electron-photon vertex in Fig.~\ref{fig:fig1}).
Finally,
when the wavefunctions for all the legs
at a vertex are known,
the amplitude for the diagram
is computed using 
an ``amplitude'' routine
(e.g.~\texttt{FFV1\_0} for the 
muon-muon-photon vertex in Fig.~\ref{fig:fig1}). 
More details on this formalism and on 
the various optimizations implemented 
within \mgamc\ are given in Ref.~\cite{bib:kiran}.

\newcommand{\tabcomp}{
\begin{table}[t]
\begin{center}
\begin{tabular}{lrrr}
\hline
\emph{Process} & \emph{LOC} & \emph{functions} &  \emph{function calls} \\
\hline
\eemumu\ & 776 & 8 & 16 \\
\ggtt\ & 839 & 10 & 22\\
\ggttg\ & 1082 & 36 & 106 \\
\ggttgg\ & 1985 & 222 & 786 \\
\hline
\end{tabular}
\end{center}
\vspace*{-2mm}
\caption{Comparison of the complexity 
of generated CUDA code 
for various physics processes,
as the number of relevant 
Feynman diagrams increases.
The columns represent the number 
of auto-generated lines (\emph{LOC}), 
the number of individual generated 
functions (\emph{functions}) 
and the number of function calls 
during one calculation (\emph{function calls}),
as several functions are called multiple times 
with different parameters. 
An additional source of complexity,
not shown in this table 
or in Fig.~\ref{fig-mecompl},
is the size
of the color matrix
required by the QCD color flow algebra,
described in Sec.~\ref{sec_mgamc}.}
\label{tab-compl}
\vspace*{-5mm}
\end{table}
}

\newcommand{\figcomp}{
\begin{figure}[b]
\vspace*{-2mm}
\begin{center}
\hspace*{-0.00\textwidth}\includegraphics[width=1.00\textwidth,clip]{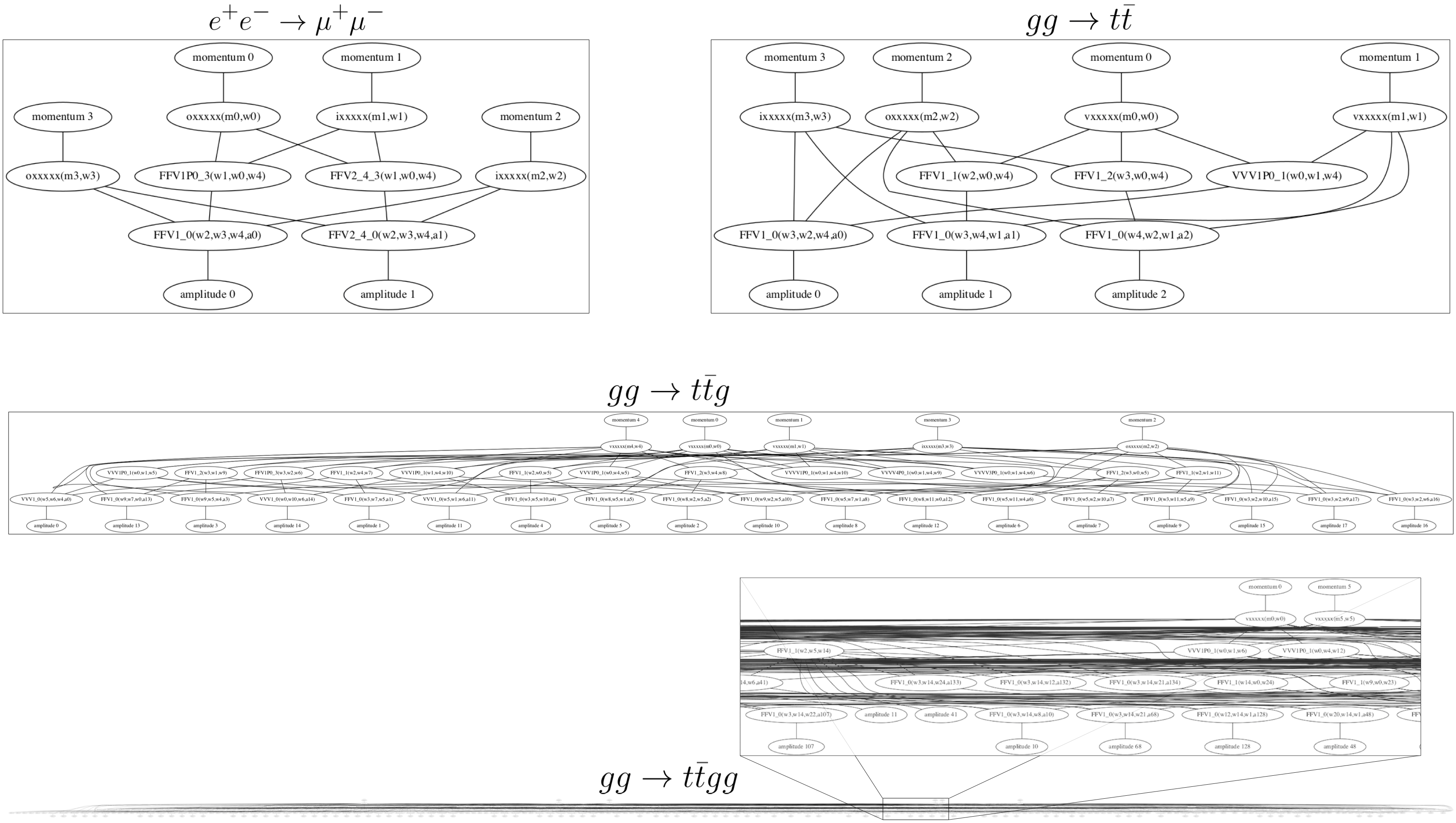}
\end{center}
\vspace*{-5mm}
\caption{visualization of the complexity 
of ME calculations 
for various physics processes.
The detailed computation algorithm 
is described in Sec.~\ref{sec_mgamc}.
The only inputs to the calculation
which vary from one event to another
are the 4-momenta of the 
initial and final state 
particles in the collision
(4 for \eemumu\ and \ggtt,
5 for \ggttg\ and 6 for \ggttgg).}
\label{fig-mecompl}
\vspace*{-4mm}
\end{figure}
}

\section{Software engineering process}
\label{sec_engproc}

\globalcounter{figure}
  
\columnratio{0.64} 
\begin{paracol}{2}
 
From an engineering perspective, 
\mgamc\ is a code generator, 
written in Python. 
Given a physics process,
it generates the corresponding
process-specific source code
in multiple implementation languages. 
The source code will then be compiled 
into an executable to produce 
the desired physics calculations. 
At the time of writing, 
\mgamc\ provides code back-ends 
to produce Fortran, C++ and Python 
source code.
Fortran is the only back-end
used in production
by the LHC experiments.

\paragraph{Iterative engineering in CUDA/C++}

As described in Sec.~\ref{sec_intro},
the primary goal of this project is to add 
one or more back-end implementation languages 
that will allow the workflow execution 
also on graphics cards from various vendors 
such  as \Nvidia, AMD or Intel. 
In this paper, we focus 
on the engineering 
for \Nvidia\ GPUs
with the C/C++ flavor of CUDA.
In the future, we 
foresee the possibility 
to support AMD GPUs
using the HIP~\cite{bib:hip} language.
Other members of our 
team are also investigating
hardware abstraction layers 
such as Alpaka~\cite{bib:alpaka}, 
Kokkos~\cite{bib:kokkos} 
and SYCL~\cite{bib:sycl},
which allow one to write 
a single implementation
that may be executed 
on various GPU (and CPU) hardware back-ends.

\switchcolumn 

\begin{figure}[h]
  \vspace*{-15mm}
  \begin{center}
    \hspace*{-5mm}
    \includegraphics[width=49mm]{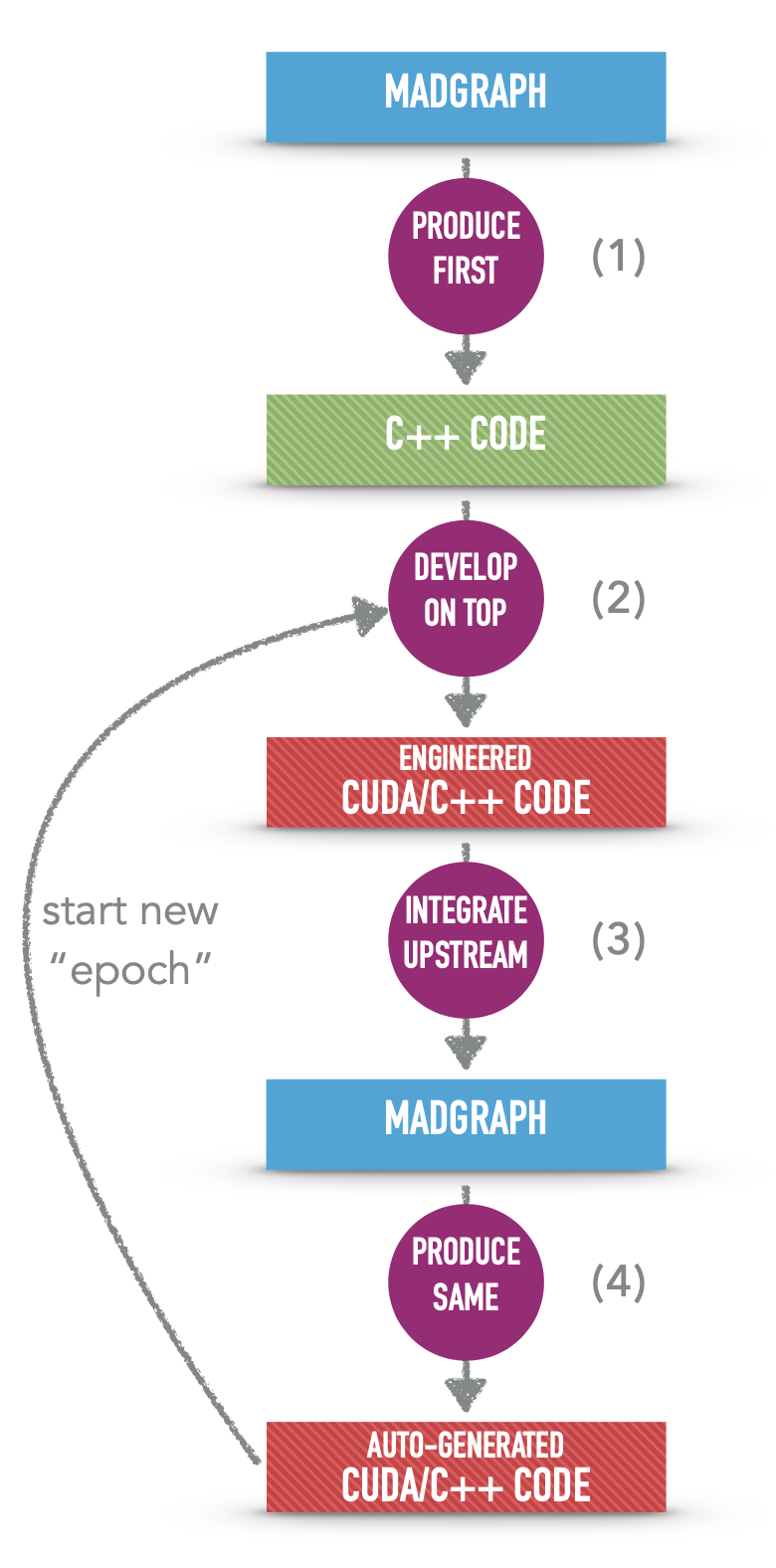}
  \end{center}
  \vspace*{-8mm}
  \caption{The overall software engineering 
  process is iterative.}
  \label{fig-sweng} 
  \vspace*{-7mm}
\end{figure}

\end{paracol}

A complementary goal for the project, which 
came about 
naturally during
our development
in the C/C++ flavor of CUDA,
is the reengineering and
performance optimization
of the C++ back-end of \mgamc\ for CPUs.
Because of the large overlap between 
the two back-ends,
we use
a single source code implementation for CUDA and C++,~where~more~than 
90\% of the code is shared,
and a few \texttt{\#ifdef}'s 
enclose
the remaining back-end-specific code.

The engineering towards 
the CUDA/C++ implementation 
is designed around a cycle of 
developing and optimizing software in CUDA/C++
and back-porting this
upstream into the 
\mgamc\ Python code 
generator. 
In  the  bootstrapping phase, 
the C++ back-end was used 
to generate the source code of the 
\eemumu\ physics process 
(see Fig.~\ref{fig-sweng}, step~1) 
in a so called ``epoch0''. 
This code was used as the basis 
to engineer a first CUDA/C++ version 
(Fig.~\ref{fig-sweng}, step~2),
which was subsequently integrated upstream 
into \mgamc\ (Fig.~\ref{fig-sweng}, step~3) 
to auto-re-generate the same CUDA/C++ code 
(Fig.~\ref{fig-sweng}, step~4), 
with a new ``epoch1'' engineering step starting. 
The newly produced code served 
as the basis towards the engineering
of a new ``epoch2'' 
in steps~2--4. 
The results we present 
in \mbox{Secs.~\ref{sec_design}--\ref{sec_results}}
are based on ``epoch2''.
At the time of writing, 
we are back-porting 
C++ vectorization
into the code generator,
which we will then use 
to produce a new ``epoch3''.
The code we develop
and additional work on abstraction layers
are hosted on github~\cite{bib:mg5-github}.
The code version
described in this paper
was also tagged and published
on Zenodo~\cite{bib:mg5-github-chep2021}.
The official \mgamc\ repository,
where our work is backported
to the Python meta-code
whenever necessary,
continues 
to be hosted
on launchpad~\cite{bib:launchpad}.

\tabcomp

Although the process \eemumu\ is 
not relevant for LHC physics, 
we chose it
as~a~starting point
for CUDA implementations
in order
to simplify 
the engineering work itself,
as it involves only a limited
number of lines of CUDA/C++ code
to work on in each epoch.
With each 
back-port of the code 
into \mgamc\ for a 
new epoch, we also 
generate source code 
for more complex QCD processes
such as 
\ggtt, \ggttg\ and \ggttgg.
This is used to validate
the code generator
and as a basis for further optimizations,
more directly relevant to LHC physics.
The increase in complexity 
of the 
auto-generated code 
for the four processes considered
is shown in Tab.~\ref{tab-compl} 
and Fig.~\ref{fig-mecompl},
for the ME calculation
part of the workflow. 

\paragraph{Simplified workflow application ---
performance optimization
and physics validation}

\figcomp

The ME calculation 
is the core component 
of \mgamc\ not 
only from a physics point of view,
but also and more importantly because
it is by far 
the largest consumer of CPU
in the overall workflow.
In a \mgamc\ Fortran executable
for \ggttgg\ events,
the ME calculation
accounts for more than 90\%
of the CPU instructions,
and this fraction further increases
for more complex LHC physics 
processes~\cite{bib:kiran}.

Our development 
so far, therefore,
has mainly focused
on optimising the
CUDA and~C++ 
software performance of 
the ME calculation within
a simplified workflow application,
rather than on 
implementing
a 
CUDA/C++ framework providing
the full functionality of \mgamc.
As described 
in Sec.~\ref{sec_design}
and Fig.~\ref{fig:anatomy},
our standalone 
application
includes 
the generation of random numbers
for a given 
number of events
using the \Nvidia\ cuRAND~\cite{bib:curand} library,
their mapping to 4-momenta
of the
particles in each collision
using a simple phase space
sampling algorithm,
RAMBO~\cite{bib:rambo},
and the ME calculation
for each event
from these 4-momenta.

The level of functional testing
that we have deemed appropriate
for this simple workflow
is 
minimal.
The 
output 
of our ``toy'' application
includes both 
performance metrics
and physics results,
which are visually inspected
when running it 
after a code change.
To validate physics 
correctness, 
the average 
of the MEs 
over all processed events
is compared to 
an expected value. 
A bitwise comparison
is possible 
if the same random number seeds
and event sample sizes are used,
as the average ME is strictly reproducible
in this case,
and even yields the same result 
in CUDA and C++
as we use a cuRAND algorithm
generating the same random sequences
on GPU and CPU;
in all other cases, 
the 
average ME result
is checked within 
its statistical error.
In general, 
the average ME
is not 
a physics relevant quantity per se,
but it is enough for
a basic check 
of the correctness 
of physics results;
in the future this will be complemented by
the proper computation of 
fiducial
cross sections. 
In addition,
functional tests
based on the 
GoogleTest~\cite{bib:googletest} API
have been developed
to compare MEs 
and particle 4-momenta 
of several individual events 
against pre-computed references,
for both CPU and GPU;
these tests are also executed
in a continuous integration in github.

\section{Software architecture design 
and implementation challenges}
\label{sec_design}

\begin{figure}[tb]
\vspace*{-3mm}
\begin{center}
\mbox{\hspace*{+0.005\textwidth}\fbox{\includegraphics[width=0.97\textwidth,clip]{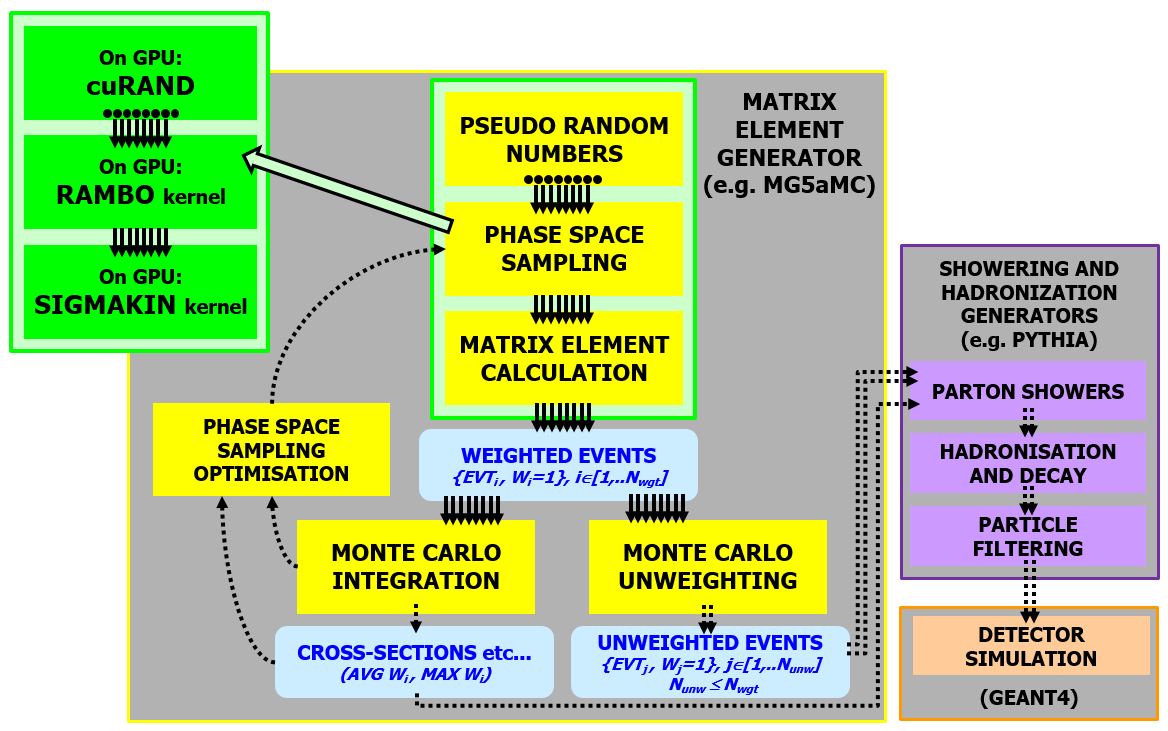}} 
}
\vspace*{-8mm}
\end{center}
\caption{Schematic representation
of the computational anatomy of
a matrix element generator. 
The three green blocks 
in the top-left corner 
represent the software components 
included in the 
standalone simplified workflow application
described in this paper.
A detailed explanation is provided in the text.}
\vspace*{-6mm}
\label{fig:anatomy}
\end{figure}

\paragraph{Component decomposition 
of Matrix Element Generators 
---
event-level data parallelism}

The very first step
in our reengineering 
of \mgamc,
and in the design of a software architecture
enabling efficient parallel processing
on both GPUs and CPUs, 
was to deconstruct
its 
computational workflow into 
separate components 
and to identify
their input and output data.
This phase of the project,
which overlapped
with the work ongoing 
in the HSF Event Generator 
WG 
for preparing
the LHCC review of HL-LHC 
computing~\cite{bib:csbs,bib:lhcctalk},
was also key to enabling
the collaboration
between people with diverse skills and backgrounds,
by translating theoretical 
physics methodologies 
into practical computational problems.

\newcommand{\Nwgt}{N_{\mathrm{wgt}}}
\newcommand{\Nunw}{N_{\mathrm{unw}}}

As described more in detail 
in Refs.~\cite{bib:csbs,bib:lhcctalk},
a Matrix Element Generator (MEG) 
like \mgamc\ is
a computer program that, starting 
from a limited set
of input configuration parameters,
is typically used to produce 
two types of outputs: 
a fiducial cross section~for 
the collision process
and 
a sample
of $\Nunw$ unweighted 
events.
Both results are obtained 
using~Monte Carlo methods,
starting from 
a larger set
of $\Nwgt$ randomly generated weighted events.
Each event 
is described 
by the 4-momenta
of all initial and final state particles,
and by an event weight,
which varies 
for weighted events,
but is a constant equal to +1
for unweighted events.
The resulting sample 
of ``parton-level'' unweighted events
and the process cross~section~are 
written out to a file
(typically using the 
Les Houches Event File, 
or LHEF~\cite{bib:lha01,bib:lhef}, format),
which
is then read back
as an input by 
a Showering and Hadronization 
Generator (SHG) 
like Pythia~\cite{bib:pythia82}.
In turn, the SHG 
adds parton showers,
hadronisation and particle decay
to each event, 
filters out events 
based on some selection criteria,
and writes 
the resulting sample 
of ``particle-level'' events
to a file. 
This file 
is then read as an input
by the detector simulation
framework Geant4~\cite{bib:geant4},
which produces yet another output
for further event processing
components down the line.
This description 
of a typical data flow 
is represented schematically
in Fig.~\ref{fig:anatomy}.
It does not cover all possible 
use cases,
some of which may be significantly 
more complex
(and involve, for instance,
jet matching and merging, or
unweighted events with negative weights),
but it is enough for 
the collision processes
that were used in this project.

The core of the MEG computation 
is the preparation of 
the sample of weighted events.
This involves
three steps,
represented by 
the green boxes
in Fig.~\ref{fig:anatomy}.
First, a set of random numbers 
is drawn for each event
using a pseudo-random number generator.
Starting from these, 
the particle 4-momenta in each event,
as well as a sampling weight,
are then constructed
by a phase space sampling algorithm.
The event weight
is finally derived
by combining the sampling weight
with the matrix element,
which is computed 
starting from these 4-momenta.

The key feature
of this computational workflow
is that the 
numerical operations
transforming input data 
into output data,
in the phase space sampling algorithm
and 
in the ME calculation,
are 
exactly the same
for all of the events
(or, at least, for all of the events 
within the same subprocess
of the overall collision process).
This makes a MEG like 
\mgamc\ an ideal candidate
to implement data parallelism
using lockstep processing on GPUs 
and vectorized code on CPUs.
As a consequence,
the engineering work in this project
relies, first and foremost,
on a software design based on
event-level data parallelism:
the basic idea is to translate 
each and every data processing step
into 
a stateless function that 
operates simultaneously 
on a basket of events,
rather than on an individual event.
This is especially important
for the ME calculation,
which 
(for all but the simplest 
physics processes)
is the computational hot-spot
of the whole workflow.
A crucial aspect of MEGs
is that 
random numbers 
are mainly used for Monte Carlo sampling,
rather than to take Monte Carlo decisions.
In other words,
conditional branching
of a stochastic nature does exist
in MEGS,
but only in a few 
well-defined situations
(such as event filtering 
due to phase space cuts,
and event unweighting),
where data stream compaction
should in principle be straight-forward
and have a computational cost
that is relatively negligible
in comparison to that 
of ME calculations.

\newcommand{\figcudatput}{
\begin{figure}[t]
\vspace*{-2mm}
\begin{center}
\hspace*{-0.01\textwidth}\includegraphics[width=1.01\textwidth,clip]{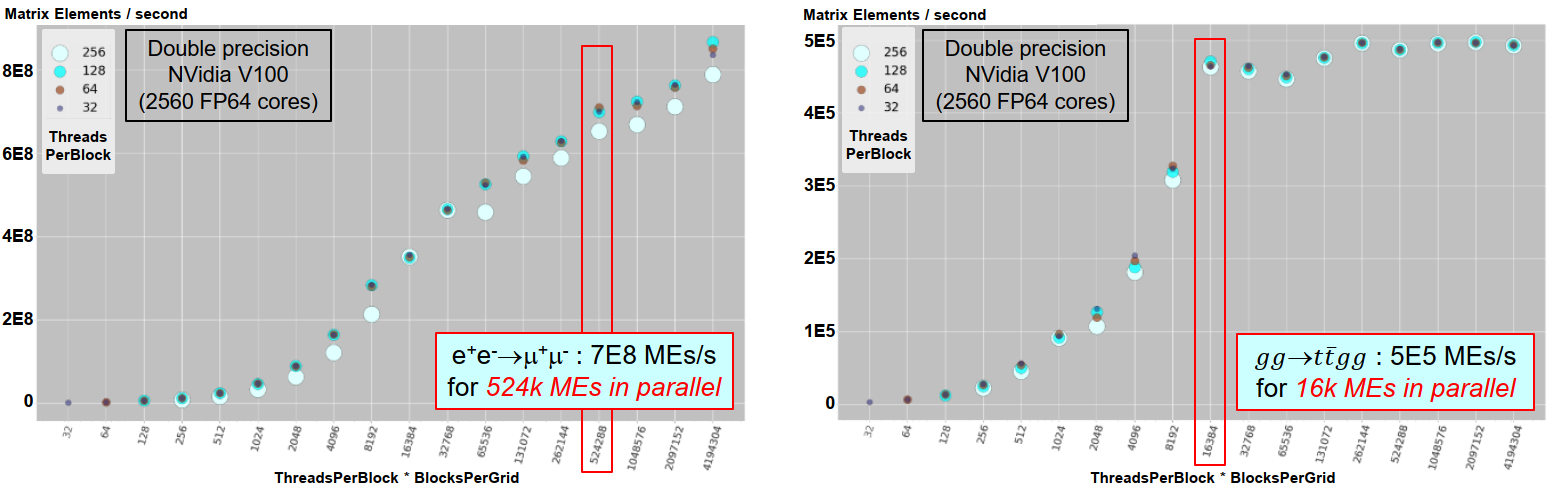}
\end{center}
\vspace*{-5.5mm}
\caption{Throughput
(MEs per second)
of the ME calculation 
in CUDA,
as a function of the 
number of threads
(i.e.~the 
number of events processed 
in one iteration),
using double precision 
floating point 
arithmetics,
for \eemumu\ (left) 
and \ggttgg\ (right).
Four 
sets of results
are plotted for 
different values
of the number of threads per block.
Throughputs are computed
as the number of MEs
divided by the sum of the
time to compute them 
on the GPU device and of
the time to copy them 
to the CPU host.
Virtual machine itscrd70
with an \Nvidia\ V100 GPU;
details of the 
software build are
given in Tab.~\ref{tab-tput}.
For illustrative purposes only,
the red boxes in the two plots
indicate two examples discussed in the text.
}
\label{fig-cudatput}
\vspace*{-6.5mm}
\end{figure}
}

\figcudatput

\paragraph{High-level design 
of software components and data structures}
In practice,
one of the most important aspects
in the software architecture 
of a data parallel system is 
the design of the relevant data structures.
As mentioned above, our 
toy 
application includes
three main components:
a random number generator,
a phase space sampler,
and a ME calculation engine.
The interface of each component
is designed to operate
on multiple events in parallel;
rather than using C++ 
vectors,
the relevant inputs and outputs 
are read from and written into
C-style arrays
with fixed dimensions,
which are properly sized
to contain 
a pre-defined number of events
that is meant to be
processed in one ``iteration''.
In our single-source code approach,
similar memory layouts 
are used for CUDA and C++,
but many specific choices
differ 
in the two implementations,
and are in any case highly configurable.
It is also worth noting that
CUDA and C++ modules 
are built in slightly different ways,
as we observed that the
CUDA implementation
is faster when
Relocatable Device Code
is disabled.

In the CUDA implementation,
each event is processed 
by one GPU~thread.~Thus,~the~number of events
in one iteration corresponds 
to the 
number of threads in a ``grid'',
i.e.~to 
the~product 
of the numbers 
of ``blocks'' per grid
and of
threads per block,
which together with the~number of iterations
are configured by three
command-line user inputs.
For \eemumu,
a~typical measurement
consists in processing 
6M events
in 12 iterations of 524k events,
partitioned in 2048 blocks of 256 threads;
in practice, one CUDA iteration 
in this case 
consists in generating 
all relevant random numbers 
(four per external particle)
for 524k events
using the cuRAND library,
and in submitting 
over a grid of 524k threads
a phase space sampling (RAMBO) kernel
and a ME
calculation (SIGMAKIN) kernel.
This specific choice of parameters
is due to the observation that
the throughput of the ME calculation
for \eemumu\ on an \Nvidia\ V100~\cite{bib:v100} GPU
only approaches its maximum plateau
when $\mathcal{O}(1\mathrm{M})$ 
threads per grid are used,
as shown in Fig.~\ref{fig-cudatput};
for comparison,
around 16k threads per grid
are enough
to reach the throughput plateau
for \ggttgg, as the ME calculation
kernel for this process
is much more computationally intensive.
A few iterations are used
to smooth out performance fluctuations,
while still requiring less than 
one second overall
for one throughput measurement
for \eemumu.

In the C++ implementation, 
data processing is also
split 
into iterations
configured using
the same three command line arguments.
While the concept of CUDA thread blocks 
has no equivalent in C++,
this is used
to configure some parameters
(e.g.~random number seeding in cuRAND)
in the same way as in CUDA,
to achieve reproducibility
of physics results on an event-by-event basis
between CUDA and C++ 
when possible.
A typical throughput measurement
for generating 6M events
over 12 iterations takes
only a few seconds
for \eemumu.

\newcommand{\figcudansys}{
\begin{figure}[t]
\vspace*{-2mm}
\begin{center}
\hspace*{-0.01\textwidth}\includegraphics[width=1.01\textwidth,clip]{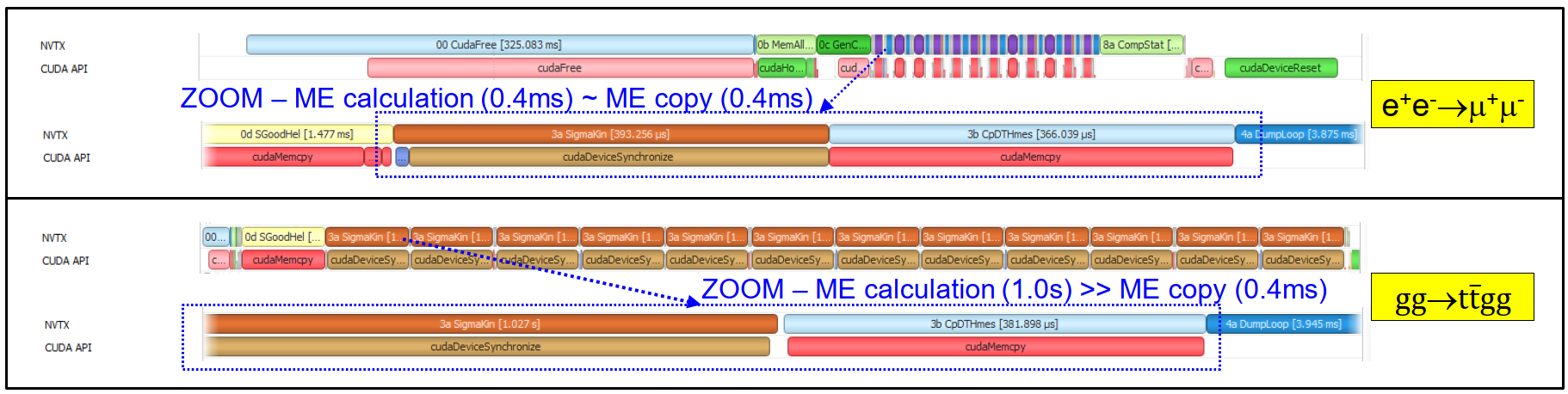}
\end{center}
\vspace*{-5mm}
\caption{Visualization
of the CUDA workflow
using \Nvidia\ Nsight 
Systems~\cite{bib:nsights},
for \eemumu\ (top) 
and \ggttgg\ (bottom).
Both sets of diagrams
refer to the execution
of the standalone executable, 
using 12 iterations with 524k events each
(2048 blocks, 256 threads per block).
The ME calculation 
takes
about 0.4ms and 1.0s 
per iteration 
for \eemumu\ and \ggttgg, respectively,
while the ME copy takes 
about 0.4ms in both cases.
Including both times
in the computation of throughputs,
this corresponds to processing rates
of about 5E5 MEs/s (5E5MEs/1.0s),
and 7E8 MEs/s (5E5MEs/0.8ms),
respectively.
Virtual machine itscrd70
with an \Nvidia\ V100 GPU;
details of the 
software build are
given in Tab.~\ref{tab-tput}.
}
\label{fig-cudansys}
\vspace*{-6mm}
\end{figure}
}

\figcudansys

Memory is allocated 
before the first iteration
and released after the last one.
Data structures are allocated
where the computations take place,
i.e.~on
the GPU device for CUDA
and on the CPU host heap for C++.
In the CUDA implementation,
4-momenta, sampling weights and MEs 
are also heap-allocated on the CPU host,
where they are copied 
after each iteration.
This data transfer 
from GPU to CPU
is slow and can add a large overhead
to the ME computation,
even if pinned host memory 
is used to speed it up.
The ratio of ME copy time
to ME computation time 
is approximately 
1:1
for \eemumu,
but this decreases
to 1:2500
for \ggttgg, as shown in Fig.~\ref{fig-cudansys}.
We therefore expect that
the overhead of 
device-to-host data transfers
should be completely negligible
for the complex processes
relevant to LHC physics.

The 
outputs 
of the cuRAND, RAMBO and SIGMAKIN steps,
namely 
random numbers,
4-momenta and weights,
are all arrays 
of floating point real numbers.
Double precision is used by default
throughout all internal calculations
in both CUDA and C++,
to be consistent with the choice made 
in the FORTRAN code 
to achieve the required physics accuracy.
At build time, 
one can also switch
to single precision:
while we initially added 
this feature to enable 
tests 
of consumer-grade GPUs,
this turned out to be very useful
to compare the 
double and single precision throughputs
of the ME calculation,
on different types 
of GPUs for data centers
or using different
levels of vectorization on CPUs,
as discussed further below.

Several 
memory layouts 
were prototyped, including
arrays of structures (AOS),
structures of arrays (SOA),
and arrays of structures of arrays (AOSOA).
In this nomenclature,
``structure'' (S) reflects the
data organization within each event,
while ``array'' (A) indicates
independent events.
All layouts are coded
as 1-dimensional C-style arrays,
but are decoded
using multi-dimensional indices.
The AOSOA layout is now the default:
data are partitioned in 
event
``pages'', where each page 
(the leading A)
represents a different subset 
of events
and is organized internally as an SOA.
For instance,
the 4-momenta AOSOA
is logically 
a four-dimensional array,
where the four 
indices run over 
the number of event pages,
the number of particles in each event,
the 4 components of 4-momenta,
and the number of events per page.

The number of events 
per page,
i.e.~the striding of the AOSOA,
is 
highly configurable
and can differ 
in the CUDA and C++ implementations
and in the random number 
and 4-momenta arrays.
As an example,
the 4-momenta AOSOA in CUDA
has 4 events per page in double precision
and 8 in single precision:
this choice ensures that
that 
a given data item 
(e.g.~the energy of the first particle)
for different events in the same page
is aligned so that it
can 
be fetched in 
one 32-byte 
transaction~\cite{bib:coalesce} on a V100 GPU.
Profiling using \Nvidia\ Nsight 
Compute~\cite{bib:nsightc}
confirms
that this choice 
achieves coalesced memory access,
i.e.~it fulfills
all data requests
({\small\texttt{l1tex\_\_t\_requests\_pipe\_lsu\_mem\_global\_op\_ld.sum}}~\cite{bib:nsightc-div})
using a minimal number
of memory transactions
({\small\texttt{l1tex\_\_t\_sectors\_pipe\_lsu\_mem\_global\_op\_ld.sum}}~\cite{bib:nsightc-div}).
With respect to AOS layouts
(which can be recovered
by using an AOSOA striding
with one event per page),
the optimal AOSOA layouts
for the 4-momenta array
reduce the number 
of memory transactions
by a factor 4 
in double precision
(or a factor 8 in single precision).
Using AOSOA's instead of AOS's,
however, only increases
the throughput of the ME calculation
by approximately 5\% to 10\%:
this is probably because,
even in the simplest \eemumu\ case,
the cost of the calculation
is much higher than that
of data access from GPU global memory.

While the AOSOA layout was initially designed
for the CUDA GPU implementation,
and specifically 
for data transfers
at the interface between
the three main components
of our simplified application,
the benefits of this type of layout,
where the variables
representing the same data item
for different events 
are adjacent to one another,
turned out to be especially important
in the C++ implementation,
and more specifically
in the internal implementation
of the SIGMAKIN component,
where this represents
an essential and unavoidable
ingredient of vectorization.
This is described in more detail 
in the following paragraph.

\newcommand{\muc}[1]{\multicolumn{2}{c|}{#1}}
\newcommand{\mur}[1]{\multirow{2}{*}{#1}}

\newcommand{\tabtput}{
\begin{table}[t] \vspace*{-3mm}
\begin{center}{
\small
\hspace*{-4mm}
\setlength\tabcolsep{5pt} 
\begin{tabular}{|c|c|c|c|c|}
\hline
\emph{CPU build name (\eemumu) 
\hfill (SIMD spec)} & 
\emph{\#doubles} & \emph{\#floats} &
\muc{\emph{MEs/sec (compute)}} \\
\cline{4-5}
\emph{\footnotesize [SIMD compiler flags] \hfill\phantom{x}} & 
\emph{per vector} & \emph{per vector} &
\emph{Double} & \emph{Float} \\
\hline
\hline
MadEvent Fortran \hfill (scalar) &
\mur{x1} & \mur{x1} &
1.50E6 & --- \\
{\footnotesize [---] \hfill\phantom{x}} &
& &
(x1.15) & \\
\hline
\hline
Standalone C++ ``none'' \hfill (scalar) &
\mur{x1} & \mur{x1} &
1.31E6 & 1.21E6 \\
{\footnotesize [---] \hfill\phantom{x}} &
& &
\bf{(x1.00)} & (x0.92) \bf{\em{[x1.00]}}\\
\hline
Standalone C++``sse4'' \hfill (128-bit SSE4.2) &
\mur{x2} & \mur{x4} &
2.52E6 & 4.50E6 \\
{\footnotesize [-march=nehalem] \hfill\phantom{x}} &
& &
(x1.9) & (x3.4) \em{[x3.7]} \\
\hline
Standalone C++ ``avx2'' \hfill (256-bit AVX2) &
\mur{x4} & \mur{x8} &
4.58E6 & 8.17E6 \\
{\footnotesize [-march=haswell] \hfill\phantom{x}} &
& &
(x3.5) & (x6.2) \em{[x6.8]} \\
\hline
Standalone C++ ``512y'' \hfill (256-bit AVX512) &
\mur{x4} & \mur{x8} &
4.91E6 & 8.84E6 \\
{\footnotesize [-march=skylake-avx512 
-mprefer-vector-width=256] \hfill\phantom{x}} &
& &
(x3.7) & (x6.7) \em{[x7.3]} \\
\hline
Standalone C++ ``512z'' \hfill (512-bit AVX512) &
\mur{x8} & \mur{x16} &
3.74E6 & 7.42E6 \\
{\footnotesize [-march=skylake-avx512 
-DMGONGPU\_PVW512] \hfill\phantom{x}} &
& &
(x2.9) & (x5.7) \em{[x6.1]} \\
\hline 
\multicolumn{5}{c}{\vspace*{-1mm}}\\
\hline
\emph{GPU build name (\eemumu)
\hfill\phantom{x}} &
\muc{\emph{MEs/sec (compute)}} &
\muc{\emph{MEs/sec (compute+DtoH)}} \\
\cline{2-5}
\emph{GPU model\hfill 
{\footnotesize (Theoretical TFlops)}}&
\emph{Double} & \emph{Float} &
\emph{Double} & \emph{Float} \\
\hline
\hline
Standalone CUDA \hfill\phantom{x} &
1.37E9 & 3.28E9 &
7.25E8 & 1.59E9 \\
NVidia V100 \hfill {\footnotesize 
(7.1/double, 14.1/float)} & 
(x1050) & (x2500) &
(x550) & (x1200) \\
\hline
Standalone CUDA \hfill\phantom{x} &
4.01E7 & 8.22E8 &
3.89E7 & 6.38E8 \\
NVidia T4 \hfill {\footnotesize 
(0.25/double, 8.1/float)} & 
(x31) & (x630) &
(x30) & (x490) \\
\hline
\end{tabular}
}\end{center}
\vspace*{-3mm}
\caption{
Preliminary results. Throughput
of the ME calculation for \eemumu,
in Fortran,~C++~and CUDA,
using double or single precision 
floating point arithmetics.
For Fortran: estimate from MATRIX1 in MadEvent.
For C++ and CUDA: measurements from the 
standalone executables, 
over 12 iterations with 524k events
(2048 blocks, 256 threads per block in CUDA).
Compilers: gcc9.2 and CUDA11.0. 
All builds use ``-O3'', and 
either ``-ffast-math'' (C++) 
or ``-use\_fast\_math'' (CUDA).
Virtual machines 
itscrd70 (Fortran, C++ and CUDA/V100 results)
and lxplus770 (CUDA/T4 results),
both using
skylake-avx512 CPUs (Intel Xeon Silver 4216) 
with 4 virtual cores.
Fortran and C++ throughputs use a single CPU core.
The rightmost CUDA throughputs 
include device-to-host copies
of all ME values.
Theoretical TFlops values 
for \Nvidia\ V100~\cite{bib:v100} 
and T4~\cite{bib:t4}
from TechPowerUp~\cite{bib:v100bis,bib:t4bis}.}
\label{tab-tput}
\vspace*{-5mm}
\end{table}
}

\paragraph{\mbox{Optimising ME calculations:
CUDA register pressure,
C++ vectorization and multithreading}}
The SIGMAKIN component,
where matrix elements 
are calculated,
is the computational hot-spot of the application.
For this reason,
this is the component
where 
most of our 
optimisation efforts
have focused so far,
and 
where we plan to 
invest more work
to obtain further performance improvements
in the future,
in both 
the CUDA and C++ implementations.

\newcommand{\ffv}{\texttt{FFV}}
As described in~Sec.~\ref{sec_mgamc},
the ME calculation 
follows the helicity amplitude formalism; 
this involves a few 
\texttt{IXXXXX} and \texttt{OXXXXX} calls 
and many calls
to ``\texttt{FFV}'' functions
like \texttt{FFV1\_0},
all of
which 
operate on particle wavefunctions
using complex number arithmetics.
To ensure 
portability between CUDA and C++,
as well as to implement vectorization in C++,
we use a custom definition of complex number types,
which is
a light wrapper
over \texttt{std::complex} for C++ and 
\texttt{thrust::complex} 
(or. optionally, \texttt{cuComplex}) for CUDA.
Double or single precision
complex numbers are used,
depending on the type chosen
for 4-momenta and MEs.

In our current CUDA implementation,
SIGMAKIN is 
a single kernel.
In practice, this 
encapsulates
all of the 
function calls in
one of the diagrams
in Fig.~\ref{fig-mecompl}.
The amount of data that this kernel 
needs to process
is therefore quite large, as it 
includes 
the wavefunctions that the 
\ffv\ functions
exchange with one another,
as well as 
the data
that each \ffv\ function 
needs internally.
This is potentially 
a problem,
especially for complex 
processes involving a large number of 
Feynman diagrams and \ffv\ functions,
because each CUDA thread
can only access a limited number 
of registers~\cite{bib:cudareg}
(e.g.~255 on a V100),
and using more data than fits in registers
leads to spilling to global memory.
Already in 
the simple \eemumu\ process,~we 
have observed that ``register pressure''
limits the performance of our code:
according to
Nsight Compute
({\small\texttt{launch\_\_registers\_per\_thread}}),
SIGMAKIN 
uses 120 registers in double 
and 48 in single precision
on a V100.
One algorithmic improvement
that has resulted in
significant increases 
in ME throughputs
(up to 2x in CUDA 
and 1.3x in C++ 
for \eemumu),
which 
for CUDA
can partly be explained
in terms of reduced 
register pressure,
has consisted in implementing
simpler versions
of the \texttt{IXXXXX} and \texttt{OXXXXX} functions,
for specific~use~cases~(e.g.\ if 
a particle mass
or transverse momentum is 0).
Another, more limited, speedup
was obtained by storing
physics model parameters
(like masses and couplings) in 
CUDA constant memory.
We have also done 
a few initial tests 
using shared memory
to hold the intermediate wavefunctions,
but a more 
comprehensive
approach is needed.
In particular,
one interesting option would consist
in splitting the large SIGMAKIN
kernel into a number of smaller,
independent kernels
(e.g.~one kernel per \ffv\ call),
possibly orchestrating the process
using CUDA Graphs~\cite{bib:cudagraphs};
this 
will certainly be an active area
of our development 
in the future.

To reduce~register pressure,
we are also considering to try
a radically different 
way to achieve
lockstep, 
by using different GPU threads
to process different helicities
for the same event.
For the time being, however,
our default approach
to achieve lockstep processing
remains event-level parallelism,
where a different event is assigned 
to each CUDA thread.
So far, in fact,
this fundamental design choice
has been a success.
Implementing this strategy
was relatively easy
and did not~require major preparatory work:
as mentioned, AOSOA layouts do provide
a benefit for CUDA 
memory access,
but strictly speaking they 
are not
essential.
Profiling our code 
using 
\Nvidia\ Nsight Compute 
indicates that the SIGMAKIN kernel
reaches a 100\% branch efficiency 
({\small\texttt{sm\_\_sass\_average\_branch\_targets\_threads\_uniform.pct}}~\cite{bib:nsightc-div}),
i.e.~does not 
exhibit 
thread divergence.
In other words, 
event-level parallelism in
our code
does 
efficiently exploit the
SIMT (Single Instruction Multiple Threads)~\cite{bib:simt}
architecture of the GPU, as
the same instructions
are 
executed in lockstep 
by the CUDA threads in a warp.

It was thus 
quite natural to also develop 
an implementation 
of SIMD (Single Instruction Multiple Data)
vectorization in our C++ code for CPUs,
similarly
based on event-level data parallelism.
In practice,
the idea is that the innermost loop
must be the loop over events,
and more specifically over
the contiguous events in one AOSOA page,
so that each 
low-level
arithmetic operation (e.g.~a sum)
on a pair of
floating point numbers
can be simultaneously
performed 
on several pairs of numbers
as a single SIMD instruction.
This has required
several disruptive changes to the code,
which had not been needed
to achieve event-level data parallelism in CUDA.
A first
specific example is that
the loop over helicities,
which in the initial CUDA implementation
was repeated in 
each event,
had to be moved outside the event loop.
This also required some modification
in the algorithm that determines
which helicity combinations are valid.
A second example is that
the interfaces of 
\texttt{IXXXXX}, \texttt{OXXXXX} 
and 
\ffv\ functions
had to be modified to 
deal with input and output data
describing several events,
instead of individual events,
and that their internal implementations
had to be changed so
that arithmetic operations
manipulate vector data
instead of scalar data.
Both of these goals were 
achieved by defining 
new vector data types,
e.g.~\texttt{fptype\_v}
for a vector of floats or doubles,
using the gcc~\cite{bib:gcccve} 
or clang~\cite{bib:clangcve}
compiler vector extensions.
A vector of complex numbers,
\texttt{cxtype\_v},
is defined as a SOA
where the real and imaginary parts
are separately stored 
in two contiguous \texttt{fptype\_v}
arrays (\texttt{RRRRIIII}),
rather than as an AOS (\texttt{RIRIRIRI}).
An additional type 
\texttt{cxtype\_sv} 
is a scalar
in CUDA
and a vector 
in C++,
so that,
similarly, a ``+'' sign
indicates a scalar sum in CUDA
and a vector sum in C++; 
the power of this approach is 
that the interface and implementation code 
of \ffv\ functions 
are formally identical 
for CUDA and C++,
which considerably simplifies
the backport 
to the code-generating Python metacode.

\tabtput

Our C++ implementation
of the ME calculation
supports different
SIMD architectures and
can currently be built
in five modes. 
These are summarized in Tab.~\ref{tab-tput},
with the corresponding 
preliminary 
throughputs achieved
on an Intel Xeon Silver 4216 CPU.
In our baseline gcc implementation,
the best results are achieved
in the ``512y'' build,
which uses AVX512
with vector widths limited 
in our source code
to 256 bits.
In this case,
where
each 
\texttt{fptype\_v} is an array of
4 doubles or 8 floats,
throughput increases 
of x3.7 and x7.3 are achieved
in double and single precision,
compared to 
the corresponding 
scalar builds.
These throughputs 
are 5\% to 10\% higher
than those achieved
using the same vector dimensions
in the AVX2 build;
disassembly using \texttt{objdump}
indicates that this is
thanks to a few additional
vector instructions
on the 256-bit \texttt{ymm} registers,
introduced in the AVX512VL extensions.
Further analyses and optimisations
are in progress on
the ``512z'' build,
based on AVX512 with default 
512-bit \texttt{zmm} 
registers, which
could theoretically
deliver a higher throughput
because it uses vectors
of 8 doubles or 16 floats,
but has so far achieved
a worse throughput than AVX2.

In our C++ implementation,
the striding 
of 4-momenta AOSOA's
is presently hardcoded to
the dimension of
\texttt{fptype\_v} arrays.
This is meant to ensure
that vectorization can 
lead to throughput increases
not only in 
floating point operations
(e.g.~because a single AVX2 instruction
executes 4 sums of doubles simultaneously),
but also in 
data access from memory
(e.g.~because a single AVX2 instruction
loads the \eplus\ energies 
for 4 different events).
Work is in progress 
on a more flexible implementation,
which will allow a 
detailed performance analysis 
of these two separate effects,
and will also simplify 
the integration
of the CUDA/C++ ME calculation
in the Fortran MadEvent infrastructure,
discussed in Sec.~\ref{sec_results}.

SIMD vectorization
and multi-core execution
are two largely independent 
concurrency mechanisms that 
should not be confused,
although they can both be 
exploited through event-level parallelism.
The above description of
the C++ implementation 
and the 
results 
in Tab.~\ref{tab-tput} 
all refer to a single CPU core.
Preliminary 
tests~\cite{bib:vchep-talk}
on a few 
multi-core CPUs
show that 
ME throughputs scale well,
up to the number of physical cores
and also in the hyper-threading regime,
if several copies of
our standalone C++ application
are executed simultaneously.
A better option is to
use multithreading (MT)
to process different 
sets of events
in separate C++ threads,
as this has the advantage
of reducing the total memory 
footprint on a node.
Our first MT implementation,
based on OpenMP~\cite{bib:omp},
is not yet delivering
a stable and satisfactory performance,
and needs further optimizations.
Eventually,
a heterogeneous ME engine
can also be developed,
where the ME computation 
runs for some events in C++
on different CPU threads,
while for other events
it is offloaded to the GPU via CUDA.
A basic prototype
of this functionality already exists,
but needs further design 
and optimisation efforts.

\section{Summary of preliminary results,
future plans and outlook}
\label{sec_results}

A summary of our preliminary results
is given in Tab.~\ref{tab-tput},
in terms of the 
throughputs
(in MEs per second)
of the 
\eemumu\ ME calculation
on two test systems.

The throughputs 
achieved 
by our C++ and CUDA ME engines
have already been partially 
discussed in the previous section.
Compared to a scalar
C++ implementation
of the ME calculation,
software vectorization
using AVX512 and 256-bit wide registers
achieves a throughput increase
by factors which
are only marginally lower
than the x4 and x8 
theoretically achievable
in double and single precision.
Neglecting the time needed 
for the device-to-host copies of MEs,
the throughputs achieved 
on a full V100 GPU
are larger than that
of scalar C++
on a single CPU core
by factors of
approximately x1000 and x2500
in double and single precision.
This large difference
is mainly due to the fact
that a V100~\cite{bib:v100} has 
twice as many FP32 cores (5120)
as FP64 cores (2560),
thus delivering 
a theoretical performance
in single precision (14.1 TFlops)
which is twice that
in double precision (7.1 TFlops).
The lower register pressure
of SIGMAKIN in single precision,
previously discussed,
also contributes an additional speedup.
The situation is very different
on a T4~\cite{bib:t4} GPU,
where we observe speedups
of approximately x30 and x600
for double and float
compared to our scalar C++ code;
the much lower performance of the T4
in single precision is 
due to the 
much larger
difference
between the maximum theoretical 
performances~\cite{bib:t4bis}
of this GPU
for FP64 (0.25 TFlops)
and FP32 (8.1 TFlops),
which are in a ratio~1:32.
We expect that newer generations 
of \Nvidia\ data center GPUs
will probably look more similar
to the V100 than the T4,
with a 1:2 ratio between
FP64 and FP32 performances:
for instance,
this is the case of 
the A100~\cite{bib:a100},
with theoretical peak performances
of 9.7 and 19.5 TFlops,
respectively.

In other words, 
both for C++ and CUDA,
even if for very different reasons,
using single precision
instead of double precision
floating point arithmetics
generally results 
in at least a factor~x2
throughput increase
(and much more on some GPUs).
This option should therefore
be carefully considered
if additional large speedups are needed.
This choice, however,
would require an important effort
to rewrite some calculations
for a better control of numeric instabilities,
followed by a careful 
validation of physics results;
even in the simple \eemumu\ baseline use case,
we have observed that 
in around one event in one million
the calculated ME takes a
Not-a-Number (NaN) value
if single precision is used,
both for CUDA and for~C++.

In addition to the 
CUDA/C++ ME throughputs
measured in our 
RAMBO-based
standalone C++ application,
Tab.~\ref{tab-tput}
includes 
a rough estimate 
of the throughput
for the ME calculation
in the production Fortran 
implementation of \mgamc,
where phase space sampling 
is driven by MadEvent.
Using similar compilation options
(in particular, fast math),
our scalar C++ implementation
on a single CPU core
seems to be 
slower than 
the corresponding Fortran version
by only 15\%.
We stress that this 
is not an
apple-to-apple comparison,
but
we believe that this
gives a reasonable order of magnitude
of the throughput gains
that could be achieved
by using our CUDA/C++ implementation
instead of Fortran
for the ME calculation.

Our current plan,
in particular,
which we have already 
discussed with the 
LHC experiments~\cite{bib:roiser21},
consists in injecting
the new CUDA/C++ implementation
of the ME calculation
into the existing, MadEvent-based,
\mgamc\ infrastructure.
This is a complex mix 
of Fortran, Python and shell scripts,
which the LHC experiments
have thoroughly validated and
routinely use
for cross section calculations and
unweighted event generation.
We therefore believe that
the fastest way to bring
our new developments to production 
consists in
reusing most of 
this existing infrastructure,
and in particular its outer shell
with its well-documented user interface,
while only replacing 
its inner core
that represents its computational bottleneck.
Amongst other things,
this strategy has the advantage
that many complex
software and physics issues,
such as event I/O to files
or the interface to PDFs 
and parton showers,
should automatically be taken care of 
by the existing \mgamc.
Many functionalities will 
however 
require further work,
such as the interface to loop libraries
required for NLO calculations.

In practice,
this 
strategy
mainly requires 
the reengineering 
of the Fortran code 
that within MadEvent
calls the ME calculation function
for a given event,
from the 4-momenta of its particles. 
In particular, 
the event loops must be modified
so that several phase 
MEs are computed in parallel
for a set of pre-prepared phase space points,
and special care must also be taken
to identify all relevant inputs and outputs
that the current ME calculation in Fortran
processes through function arguments
or, more often, Fortran common blocks.
Care must also be taken 
in the way 4-momenta data is passed
from Fortran to C++,
although the overhead 
of suboptimal data loading
is expected to become negligible
as the complexity 
of the ME calculation increases.
Work in this area
has already started 
and is making rapid progress.

It should be stressed that 
the throughput increase factors
quoted above 
do not refer to a full software workflow
for calculating cross sections
or generating weighted or unweighted events,
but only to the component
responsible for the ME calculation.
Amdahl's law~\cite{bib:amdahl,bib:amdahl1967}
puts stringent constraints 
on the achievable increase
in the throughput 
of the full software workload,
if a significant fraction of it
cannot be parallelized.
Taking into account,
however, that the ME calculation
often accounts for much more 
than 90\% of the total CPU time,
overall speedups 
by a factor~3x or more
on a CPU through vectorization
seem to be within reach.

While our focus is 
shifting
to the integration of Fortran and CUDA/C++
as described above,
we will keep using 
the simplified CUDA/C++ standalone executable
as our workhorse for all ME optimisations.
We also plan to continue prototyping 
some functional enhancements
of this application,
albeit at a much lower priority.
This ranges from the implementation
of proper cross section calculations
and event unweighting,
to improvements in phase space sampling
(through the vectorization of RAMBO,
the reimplementation of MadEvent in CUDA/C++
or the use of external packages 
such as VegasFlow~\cite{bib:vegasflow}),
to the integration of
parton distribution functions
(possibly through external packages
such as PDFFlow~\cite{bib:pdfflow}).
In this context,
we also closely follow
the related developments
by the MadFlow 
team~\cite{bib:madflow-talk,bib:madflow}.

Finally, we believe
that our implementations
of ME calculations
in CUDA (and eventually other languages) for GPUs
and in C++ for vector CPUs
represent very useful 
software workloads
for the benchmarking
of computing resources
used by the HEP experiments,
in an increasingly heterogeneous environment.
In this context,
we are collaborating
with the HEPiX benchmarking working group,
with the goal of eventually providing
a \mgamc-based 
containerized standalone application
to be integrated
in the HEP-Benchmarks 
suite~\cite{bib:bmkav,bib:bmkmfm}.
We also plan to port our code
to new architectures,
such as 
AMD and Intel GPUs
but also 
non-x86 CPUs
such as ARM and Power9,
also in view of the 
use of 
the software at HPC facilities.

\section{Conclusions}
\label{sec_conclude}
\label{sec:concl}

In this paper,
we have presented
an ongoing project
aiming at 
reengineering the 
\mgamc\ physics event generator,
primarily to develop 
a new CUDA back-end for GPUs,
but also 
to optimize the performance
of its C++ back-end on CPUs.
We have described its motivation,
its iterative engineering process
involving code-generating metacode, 
its software architecture
based on event-level data parallelism
to efficiently exploit SIMT on GPUs
and SIMD on CPUs,
as well as some 
current challenges
and future directions.
We have also summarized
the preliminary results
that we have 
presented at the vCHEP2021
conference~\cite{bib:vchep-talk},
providing~an 
update 
over previous 
preliminary results
shown
in the past~\cite{bib:roiser20}.
Our first results are very encouraging,
showing that speedups 
of the ME calculation
by more than a~factor~3~on~CPUs
are possible through SIMD vectorization,
and by three orders of magnitude
on a typical data center GPU.
We are now focusing
on a strategy 
to bring these improvements 
to a production release of the code
usable by the LHC experiments
in a relatively short time,
by injecting our new CUDA/C++ component
into the existing Fortran-based
infrastructure of \mgamc.
This is work in progress,
about which 
further updates
will be presented
in later publications.

\section*{Acknowledgements}

It is a pleasure to 
thank our colleagues 
in the 
madgraph4gpu
project,
Tyler Burch, Taylor Childers, Smita Darmora, 
Laurence Field, Walter Hopkins, Josh McFayden, 
Vince Pascuzzi, Markus Schulz, David Smith
and Carl Vuosalo,
for our fruitful 
collaboration, 
and Andy Reepschlaeger and Taran Singhania 
also for their contributions to 
the infrastructure for
the analysis of performance metrics.
We are 
grateful 
to Sebastien Ponce,
Hadrien Grasland and Marco Clemencic
for many 
suggestions
about SIMD vectorization,
and to Ricardo Rocha 
and 
CERN IT-CM
for their help in setting up
our GPU build nodes.
We thank Domenico Giordano and 
the 
HEPIX benchmarking working group,
as well as Helge Meinhard
and the WLCG benchmarking task force,
for stimulating 
discussions 
on benchmarking
of GPUs and heterogeneous systems.
Useful discussions 
with Stefano Carrazza and the MadFlow team,
as well as with
Ingvild Brevik Hoegstoeyl and her colleagues 
in NTNU and OpenLab,
are gratefully acknowledged.
We thank the authors
of the \mgamc\ software
for their continued support.

We are indebted to
the organizers of the 
Sheffield GPU Hackathon~\cite{bib:shef}.
We particularly wish to thank
Andreas Herten, Peter Heywood 
and Mateusz Malenta
for the support they provided 
as our mentors 
and their 
many useful suggestions
about our CUDA development.

\renewcommand{\baselinestretch}{0.958}

\end{document}